%% file: bsg_sd.tex
\documentstyle[12pt,axodraw,epsfig,cite]{article}
\def\be{\begin{equation}}
\def\bea{\begin{eqnarray}}
\def\ee{\end{equation}}
\def\eea{\end{eqnarray}}
                              \def\barr{\begin{array}}
                              \def\earr{\end{array}}
\def\dis{\displaystyle}
\def\etal{{\em et al.}}

                              \def\gev{\: \rm GeV}

\def\gappeq{\mathrel{\rlap {\raise.5ex\hbox{$>$}}
            {\lower.5ex\hbox{$\sim$}}}}
\def\lappeq{\mathrel{\rlap{\raise.5ex\hbox{$<$}}
            {\lower.5ex\hbox{$\sim$}}}}
\def\ra{\rightarrow}

\def\slashiii#1{\setbox0=\hbox{$#1$}#1\hskip-\wd0\hbox to\wd0{\hss\sl/\/\hss}}

\begin{document}                                                              
%===================================================================%
\begin{flushright}
MRI-P-000801\\
{\large \tt hep-ph/0008165}
\end{flushright}

\vspace*{2ex}

\begin{center}
{\Large\bf $b \to s \gamma$ confronts $B$-violating scalar couplings: 
     $R$-parity violating supersymmetry or diquarks}

\vskip 15pt

{\sf Debrupa Chakraverty$^{1}$ and Debajyoti Choudhury${^2}$}

{\footnotesize Mehta Research Institute, 
Chhatnag Road, Jhusi, Allahabad 211 019, India. \\
E-mail: $^1$rupa@mri.ernet.in, $^2$debchou@mri.ernet.in}\\

\vskip 15pt
{\Large\bf Abstract}
\end{center}

We investigate the possible role that baryon number violating 
Yukawa interactions  may take in the inclusive decay $ B \to X_s \gamma$. 
The constraints, derived using the experimental results of 
the CLEO collaboration,
 turn out, in many cases, to be more stringent than the existing bounds.

\vskip 20pt

%PACS numbers: 12.60.Jv, 14.80.Ly, 13.88.+e

\vskip 20pt
\hspace*{0.65cm}

\def\baselinestretch{1.0}

%===================================================================%
 
The Standard Model (SM) of the strong, weak and electromagnetic interactions
is in very good agreement with almost all present experimental data, 
even though a few important predictions have not yet been tested. 
Still, most physicists would readily admit that the SM cannot be the final 
theory, both on aesthetic grounds as well as on account of 
certain well-founded 
technical objections.
As a result, numerous attempts have been and are being
made in the quest of a more fundamental theory.
Experimentally, there have been two main strategies to probe new physics. 
On the one hand, we attempt to directly produce, and observe, 
new particles at high energy colliders. 
On the other, we look for virtual effects of such particles 
and/or interactions in various low and intermediate energy processes.
The decay $b \to s \gamma$ 
is an excellent candidate for the latter 
option~\cite{inami_lim,grin,drei,berto,peter,gg,rizzo,kundu,arc,
gab1,gab2,besmer,aoki,gab3}. Experimentally,
the branching ratio for the inclusive decay $B \to X_s \gamma$ have 
been measured by CLEO \cite{cleo} and ALEPH \cite{aleph} to be
\be
BR(B\to X_s \gamma) = 
\left\{ 
   \barr{rcl} 
   (3.15\pm 0.93) \times 10^{-4} & \quad & ({\rm CLEO}) \\[2ex]
   (3.11\pm 1.52) \times 10^{-4} & &       ({\rm ALEPH}) \ .
   \earr
\right.
\ee
The above are in good agreement with each other and 
with the SM prediction~\cite{kagan} of $BR(B\to X_s \gamma)
 = (3.29\pm0.33) \times 10^{-4}$. 
While a small window for the contribution of new physics 
 does remain, this agreement can obviously be used to constrain 
deviations from the SM.

In this paper, we investigate the influence that a scalar diquark
may have on the above decay\footnote{A brief discussion on the sensitivity
 of the branching ratio $B \to X_s \gamma$ to scalar diquark-top contribution 
has been presented in \cite{kagan}.}. Diquarks abound in many grand 
unified theories (with or without supersymmetry) and 
even in composite models~\cite{diquarks}. 
While vector diquarks are constrained to be 
superheavy\footnote{We do not consider the case of non-gauged 
		vector diquarks as such theories are nonrenormalizable.}
(with masses of the scale of breaking of the additional
gauge symmetry), no such 
restrictions apply to the masses of scalar diquarks. Consequently, such  
particles can be as light as the electroweak scale.
 For example, a diquark 
like behaviour can be found even in a  low energy theory like 
the Minimal Supersymmetric Standard Model (MSSM), albeit in the version 
with broken $R$-parity. 

%%%%%%
\input{table.qnos}
%%%%%%
A generic diquark is a scalar or vector particle that couples to a 
quark current with a net baryon number $B = \pm 2/ 3$. Clearly,
this may transform as either a $SU(3)_c$ triplet or sextet. 
Concentrating on the scalars (for reasons mentioned above), the 
generic Yukawa term in the Lagrangian can be expressed as 
\be
{\cal L}^{(A)}_Y= h_{ij}^{(A)} {\bar q}_i^c P_{L,R} q_j \Phi_A + h.c.,
\ee
where $i,j$ denote quark flavours, $A$ denotes the diquark type
and $P_{L,R}$ reflect the quark chirality.
Standard Model gauge invariance demands that a scalar diquark
transform either as a triplet or as a singlet under $SU(2)_L$
and that it have a $U(1)$ hypercharge 
$\vert Y\vert={2\over 3},~{4\over 3},~ {8\over 3}$. The full list 
of quantum numbers is presented in Table~\ref{tab:qnos}.
It is clear that the couplings 
 $h^{(1)}_{ij}$, $h^{(4)}_{ij}$, 
$h^{(5)}_{ij}$ and $h^{(7)}_{ij}$ must be symmetric under the exchange of 
$i$ and $j$ while $h^{(2)}_{ij}$, $h^{(3)}_{ij}$, 
$h^{(6)}_{ij}$ and $h^{(8)}_{ij}$  must be antisymmetric. For the 
other two, {\em viz.} $\tilde h^{(3)}_{ij}$ and $\tilde h^{(4)}_{ij}$, there 
is no particular symmetry property. Note that the quantum numbers 
of $\Phi_{2, 4, 6}$ allow them to couple to a leptoquark 
({\em i.e.} a quark-lepton) current as well. This implies that these 
particular diquarks could also mediate lepton-number ($L$) violating 
processes. Clearly, such leptoquark couplings need to be 
suppressed severely so as to prevent rapid proton decay. 

We make a brief interlude here to discuss the MSSM~\cite{mssm}. 
Whereas $B$ and $L$ are (accidentally) preserved in the SM (at least 
in the perturbative context), it is not so within the MSSM. 
Supersymmetry and gauge invariance, together with the field content, 
allow terms in the superpotential that violate either $B$ or
$L$~\cite{wsy}. Catastrophic rates for proton decay can be avoided though 
by imposing a global $Z_2$ symmetry~\cite{fayet} 
under which the quark and lepton superfields change by a sign,  while the 
Higgs superfields remain invariant. Representible as 
$R \equiv (-1)^{3 B - L + 2 S}$, where $S$ is the spin of a field, 
this ``$R$-parity'' is positive for the SM fields and negative for 
all the supersymmetric partners. However, while this symmetry 
is useful in preventing phenomenologically unacceptable terms, it 
has no theoretical foundation and is entirely {\it ad hoc} in nature. 
Hence, it is of interest to examine the consequences of violating this 
symmetry, not in the least because it plays a crucial role in 
the search for supersymmetry. In our study, we shall restrict 
ourselves to the case where only the $B$-violating terms are 
non-zero. Such scenarios can be motivated from a class of supersymmetric 
GUTs as well~\cite{Bviol}. The corresponding terms in the superpotential 
can be parametrized as 
\be
	W_{R\!\!\!/} = 	\lambda''_{ijk} \bar{U}^i_R  \bar{D}^j_R \bar{D}^k_R
	\label{superpot}
\ee
where $U^i_R$ and $D^i_R$ denote the right-handed up-quark 
and down-quark superfields respectively.  The couplings 
$\lambda''_{ijk}$ are antisymmetric under the exchange of the 
last two indices. The corresponding Lagrangian can then 
 be written in terms of the component fields as:
\be
{\cal L}_{R\!\!\!/}
= \lambda''_{ijk} \left(u^c_i d^c_j \tilde{d}^*_k +
u^c_i \tilde{d}^*_j d^c_k + \tilde{u}^*_i d^c_j d^c_k\right) + {\rm h.c.}
\label{lagrp}
\ee
Thus, a single term in the superpotential corresponds to {\em three different} 
diquark interactions, namely two of type $\tilde h^{(4)}_{ij}$  and one of 
type $h^{(8)}_{ij}$. 

The best direct bound on diquark type couplings comes from the 
analysis of dijet events by the CDF collaboration~\cite{CDF_dijet}. 
Considering the process $q_i q_j \ra \Phi_A \ra q_i q_j$, an 
exclusion curve in the ($m_{\Phi_A}, h^{(A)}_{ij}$) plane can 
be obtained from this data. Two points need to be noted though. 
At a $p \bar p$ collider like the  Tevatron, 
the $u u $ and $d d $ fluxes tend to be small
and hence the bounds are relatively weak. 
This is even more true for quarks of the second or third 
generation (which are relevant for the couplings that 
we are interested in). Secondly, 
such an analysis needs to make assumptions regarding the 
branching fraction of $\Phi_A$ into quark pairs, a point that 
is of particular importance in the context of 
$R$-parity violating supersymmetric models. 

There also exist some constraints derived from low energy processes. 
Third generation couplings, for example, can be constrained 
from the precision electroweak data at LEP \cite{b_c_s} or, to 
an extent, by demanding perturbative unitarity to a high 
scale~\cite{biswa}. 
Couplings involving the first two generations, on the other 
hand, are constrained\footnote{Although many of these analyses have 
		been done for the case of $R$-parity violating 
		models, clearly similar bounds would also apply 
		to nonsupersymmetric diquark couplings as well.}
by the non-observance of neutron-antineutron 
oscillations or from an analysis of 
rare nucleon and meson decays~\cite{sher,probir}. 
While many of these individual bounds are weak, 
certain of their {\em products} are much more severely constrained 
by the data on neutral meson mixing and $CP$--violation
in the $K$--sector~\cite{barbieri}. It is our aim, in this article, 
to derive analogous but stronger bounds.

%%%%%%%
\input{fig_bsphot}
%%%%%%%
Within the SM, the quark level transition $b\to s \gamma$ is mediated,
at the lowest order, by electromagnetic penguin diagrams shown in 
Fig.~\ref{fig:feyn_bsphot}($a$--$d$). While only the 
top-quark diagrams have been shown, for consistency's sake,
other charge $2/3$ quarks should also be included. However, these 
contributions are negligible on two counts: 
($i$) the small mixing angles and ($ii$) the corresponding loop 
integrals being suppressed to a great extent due to the smallness of the 
light quark masses. The matrix element for this
process is then governed by the dipole operator:
\be
-{{4 G_F} \over \sqrt{2}} V^*_{ts}V_{tb}({e\over {32 \pi^2}}) 
	C_7^{SM}(m_W){\bar s}\sigma_{\mu \nu}
 F^{\mu \nu}[m_b(1+\gamma_5)]b.
\ee
The QCD corrections to this process are calculated via an 
operator product expansion based on the effective Hamiltonian
\be
H_{\rm eff}=-{{4 G_F} \over \sqrt{2}} V_{ts}^*V_{tb}\sum_{i=1}^8 
			C_i(\mu) O_i(\mu)
\ee
which is then evolved from the electroweak scale down to $\mu=\mu_b$ 
through renormalisation group (RG) equations. A large correction 
owes itself to  the chromomagnetic operator $b\to s G$ ($G$ being a gluon)
\be
-{{4 G_F} \over \sqrt{2}} V^*_{ts}V_{tb}({g_s\over {32 \pi^2}}) 
        C_8^{SM}(m_W){\bar s}_\alpha 
\sigma_{\mu \nu}
 G^{\mu \nu}_{\alpha \beta}[m_b(1+\gamma_5)]b_{\beta},
\ee
which arises from the diagrams of Fig.~\ref{fig:feyn_bsglue}(a-c)), 
The Wilson coefficients $C_7^{SM}(m_W)$ and $C_8^{SM}(m_W)$ 
can be evaluated perturbatively~\cite{grinw,grig,buras} at the $W$ 
scale where the matching conditions are imposed. 
The explicit expressions are
\be
\barr{rcl}
C_7^{SM}(m_W) & = & \dis 
	x\left[  {{7 - 5 x - 8 x^2} \over {24 (x-1)^3}} 
             + {{x (3 x -2)} \over {4 (x-1)^4}} \ln{x} \right] \ ,
	\\
 & & \\
C_8^{SM}(m_W) & = & \dis 
	x \left[{ {2 + 5 x - x^2} \over { 8 (x -1)^3}} 
                        - {{ 3 x} \over {4 (x -1)^4}} \ln{x}\right] \ ,
\earr
\ee
where $x= m_t^2/m_W^2$.  

%%%%%%
\input{fig_bsgl}
%%%%%%

The leading order results for the Wilson coefficients at $\mu_b$, the 
$B$-meson scale, is given by
\be
C_7(\mu_b)=\eta^{16/23}\left [ C_7^{SM}(m_W) - {8 \over 3} C_8^{SM}(m_W) 
[1 - \eta^{-2/23}]
+{232 \over 513} [1 -\eta^{-19/23}] \right ]
\label{bsg2}\\
\ee
with $\eta \equiv \alpha_s(m_W) / \alpha_s(\mu_b)$,  
calculated using the leading $\mu$  dependence of $\alpha_s$, 
and the present world average value of the strong coupling constant
{\em viz.} $ \alpha_s(m_Z) = 0.118 \pm 0.005$. 
To this order, then, 
\be
\Gamma(b \to s \gamma) = {{\alpha G_F^2 m_b^5} \over { 32 \pi^4}} 
		\left| V_{tb} V_{ts}^* C_7^{SM}(\mu_b) \right|^2,
\ee
where $\alpha$ is the fine structure constant.
As the above decay rate suffers from large uncertainties due to 
$m_b$ and the CKM matrix elements, it is prudent to 
normalise it against the measured semileptonic decay 
rate of the $b$ quark
\be
\Gamma(b\to c e {\bar \nu_e}) 
        = {G_F^2 m_b^5 \over {192 \pi^3}} \kappa(z) g(z)
 \vert V_{cb} \vert^2,
\ee
where $z= m_c^2 / m_b^2$ and
\[
g(z) \equiv  1 - 8 z + 8 z^2 - z^4 - 12 z^2 \ln{z} 
\]
is the phase space factor. The 
analytic expression for $\kappa(z)$, 
the one loop QCD correction to the semileptonic decay, 
can be found in Ref.\cite{misiak}. 
The explicit dependence on $m_b^5$ is thus removed, while 
the ratio of the CKM elements in the scaled decay rate {\em viz.},
\be
\left| {{V^*_{ts} V_{tb}} \over V_{cb}} \right| = 0.976 \pm 0.010.
\ee
is much better known than the individual elements. 

An updated next to leading order (NLO) analysis \cite{misiak} 
of the $B \to X_s \gamma$ branching ratio
with QED corrections has been presented in Ref.\cite{kagan}. 
Incorporating both the NLO QCD and  
the resummed QED corrections, the Wilson coefficient $C_7^{\rm eff}(\mu_b)$
in SM can be expanded as
\be
C_7^{\rm eff}(\mu_b) = C_7(\mu_b) + {{\alpha_s(\mu_b)} \over {4 \pi}}
        C_7^{(1)}(\mu_b) +
 {\alpha \over {\alpha_s(\mu_b)}} C_7^{(em)}(\mu_b) \ .
\ee
For brevity's sake, we do not give here the expressions for 
$C_7^{(1)}(\mu_b)$ and $ C_7^{(em)}(\mu_b)$ as these can be found in 
Ref.\cite{misiak} and Ref.\cite{kagan} respectively.
The inclusion of the NLO and QED corrections in the 
$b \to s \gamma$  decay rate 
has significantly reduced the large uncertainty present in the 
previous LO calculation. From the quark level
$b \to s \gamma$ decay rate, it is possible to infer the $B$ 
meson inclusive branching ratio 
$BR(B \to X_s \gamma)$ by including the nonperturbative $1/m_b$ 
and $1/m_c$ corrections. These bound state  corrections also 
have been taken into account in Ref.\cite{kagan}. 

Having delineated the formalism, it now remains to calculate 
the additional contributions due to the possible presence of 
non-zero diquark couplings. At the one-loop level, the only 
new contributions to $b \ra s \gamma$ and $b \ra s G$ 
arise from the diagrams of Fig.~\ref{fig:feyn_bsphot}($e$--$h$) and 
Fig.~\ref{fig:feyn_bsglue}($d$--$g$) 
respectively\footnote{Clearly, to this order, none of 
	$\Phi_{5, 6}$ can mediate either of these processes and hence 
	we shall not consider such fields any further.}.
In the generic case, apart from modifications in the coefficients 
of $O_{7,8}$, two additional operators arise. Denoted as
\be
\barr{rcl}
{\tilde O}_7 & = & \dis {e\over {32 \pi^2}} \: \bar s \sigma_{\mu \nu}
 F^{\mu \nu}[m_b(1-\gamma_5)]b
         \\[2ex]
{\tilde O}_8 & = & \dis {g_s\over {32 \pi^2}} \: {\bar s}_\alpha 
\sigma_{\mu \nu}
 G^{\mu \nu}_{\alpha \beta}[m_b(1-\gamma_5)]b_{\beta} \ ,
\earr
\ee
they differ from their SM counterparts in their chirality 
structure\footnote{In all of these four operators, contributions proportional
	 to the strange quark mass have been neglected.}. 

To keep the analysis simple, 
we shall assume that only one diquark multiplet is 
light and that all the fields within a multiplet are 
degenerate\footnote{Large splittings within a multiplet is disfavoured 
	by the electroweak precision data.}. 
With this simplifying assumption, the new contributions, at the electroweak 
scale, are given by
\be
\barr{rcl}
 C^{D}_7(m_W) & = & \dis \frac{N_c}{{\cal A}}
\left [ \left\{ Q_{\Phi} F_1(y) + Q_t F_3(y)
        \right\} l_b l_s^* 
     + {m_t\over m_b}
        \left\{ Q_t F_4(y) - Q_{\Phi} F_2(y)
	\right\} r_b l_s^*
 \right] \ ,
		\\[3ex]
{\tilde C}^{D}_7(m_W) & = & \dis \frac{N_c}{{\cal A}}
 \left[\left\{ Q_{\Phi} F_1(y) + Q_t F_3(y)
	\right\}  r_b r_s^* 
     + {m_t\over m_b}
 	\left\{  Q_t F_4(y) - Q_{\Phi} F_2(y)
	\right\}  l_b r_s^*
 \right] \ ,
                \\[3ex]
C^{D}_8(m_W) & = & \dis {\cal A}^{-1}
 \left[ \left\{ C_{\Phi}  F_1(y) + C_t F_3(y)
	\right\}  l_b l_s^* 
     + {m_t\over m_b}
 	\left\{  C_t F_4(y) - C_{\Phi} F_2(y)
	\right\}  r_b l_s^*
 \right] \ ,
                \\[3ex]
{\tilde C}^{D}_8(m_W) & = & \dis {\cal A}^{-1}
 \left[ \left\{ C_{\Phi} F_1(y) + C_t F_3(y)
	\right\}  r_b r_s^* 
     + {m_t\over m_b}
 	\left\{  C_t F_4(y) - C_{\Phi} F_2(y)
	\right\}  l_b r_s^*
 \right] \ ,
          \\[3ex]
{\cal A} & \equiv & - 4 \sqrt{2} \: G_F V_{ts}^* V_{tb} m_\Phi^2 \ ,
\earr
	\label{Wilsons}
\ee
where
\be
\barr{rcl}
F_1(y) & = & \dis {1 \over {12 ( y - 1)^4}} 
	\left[ 6 y^2 \ln{y} - 2 y^3 - 3 y^2 + 6 y -1 \right] \ , \\
F_2(y) & = & \dis {1 \over {2 ( y - 1)^3}} \left[ 1 -y^2 + 2 y \ln{y}\right]
		 \ , \\
F_3(y) & = & \dis {1 \over {12 ( y - 1)^4}} 
	\left[  2 + 3 y - 6 y^2 + y^3 + 6 y \ln{y} \right] \ , \\
F_4(y) & = & \dis {1 \over {2 ( 1 - y)^3}} 
	\left[  3 - 4y + y^2 + 2 \ln{y} \right] \ ,
\earr
	\label{the_f's}
\ee
with $y= m_t^2/m_\Phi^2$. The color factor $N_c$ is 
$-1$ and $2$ for triplet and $\bar 6$ scalar respectively. 
$Q_t$ and $Q_\Phi$ are the charges of
top quark and the diquark respectively. 
The color factors $C_t$ and $C_\Phi$---
for the diagrams of 
Fig.\ref{fig:feyn_bsglue}$f$ and Fig.\ref{fig:feyn_bsglue}$g$---are given
in Table~\ref{tab:consts}. 

%%%%%%
\input{table.consts}
%%%%%%
Note that, once again, we consider only such contributions, as 
involve the top quark. As is easy to ascertain from eq.(\ref{Wilsons}),
for other quarks in the loop, the corresponding integrals are 
too small to be of any consequence. Thus, any coupling to the 
diquarks $\Phi_{7, 8}$, for example, would not be constrained to an
appreciable degree by radiative $b$ decays.
In Table~\ref{tab:consts}, we also display the relevant 
chiral Yukawa couplings (to the $b$- and $s$-quarks) 
for different choices of the diquark.

In estimating the effects of scalar diqaurk couplings, it is useful to 
consider the ratios~\cite{kagan}
$\xi_{7, 8}$ with 
\be 
\xi_{7}  \equiv 1 + \frac{C_7^{D}(m_W)}{C_7^{SM}(m_W)}, 
\ee
and similarly for $\xi_8$. 
For the new operators $\tilde O_7$ and $\tilde O_8$, we define,
\be
\tilde \xi_7 = \frac{\tilde C_7^{D} (m_W)} {C_7^{SM}(m_W)}
\ee
and $\tilde \xi_8$ in an analogous fashion.
With these definitions, the $B \to X_s \gamma$ branching ratio can 
be written as,
\bea
BR(B \to X_s \gamma) & = &  
	B_{22}(\delta) + (\xi_7^2 +  {\tilde \xi_7}^2) \: B_{77}(\delta)  
       + (  \xi_8^2 +  {\tilde \xi_8}^2) \: B_{88}(\delta) 
 + \xi_7 \: B_{27}(\delta) \nonumber\\
&  + & \xi_8 \: B_{28}(\delta) 
      + (\xi_7 \xi_8 + {\tilde \xi_7}{\tilde \xi_8}) \: B_{78}(\delta).
\label{bs1}
\eea
 In a parton level analysis, the photon  would be  monochromatic, with 
$E_\gamma = E_{\gamma}^{max} = m_b/2$. However, 
once the gluon Bremsstrahlung 
contribution is included, the photon spectrum becomes nontrivial
and, for experimental purposes, one needs to make 
an explicit demand on the photon 
energy, namely 
\be
E_\gamma > (1 -\delta) E_{\gamma}^{max},
\ee
where $\delta$ is the fraction of the spectrum above the cut.
The values of $B_{ij}(\delta)$ are listed in ref.\cite{kagan} for different 
choices of the renormalisation scale $\mu_b$ 
and the cut off parameter on the photon energy $\delta$. 
As is well known, some ambiguities exist 
in the choice of $\mu_b$ which should, typically, lie in the 
region $m_b/2$ to $2 m_b$.
For our analysis, we used $\mu_b = m_b$ and $\delta=0.9$. 
We have checked that other values of $\delta$ do not change 
the bound significantly. 

Since the new physics becomes operative only above the electroweak scale, 
the additional contributions to the operators $O_7$ and $O_8$ will 
only serve to change the Wilson coefficients at $m_W$.
Of course, the additional operators ${\tilde O_7}$ and ${\tilde O_8}$ 
would influence the RG equations for $C_7$ and $C_8$ as well. 
However, since we are primarily interested in small $C_{7,8}^{D}(m_W)$,
it is safe to neglect any term in the RG equations involving these 
coefficients.

%%%%%%%%%%%%%%%%%%
\begin{figure}[h]
\centerline{
\epsfxsize=6.5cm\epsfysize=6.0cm
                     \epsfbox{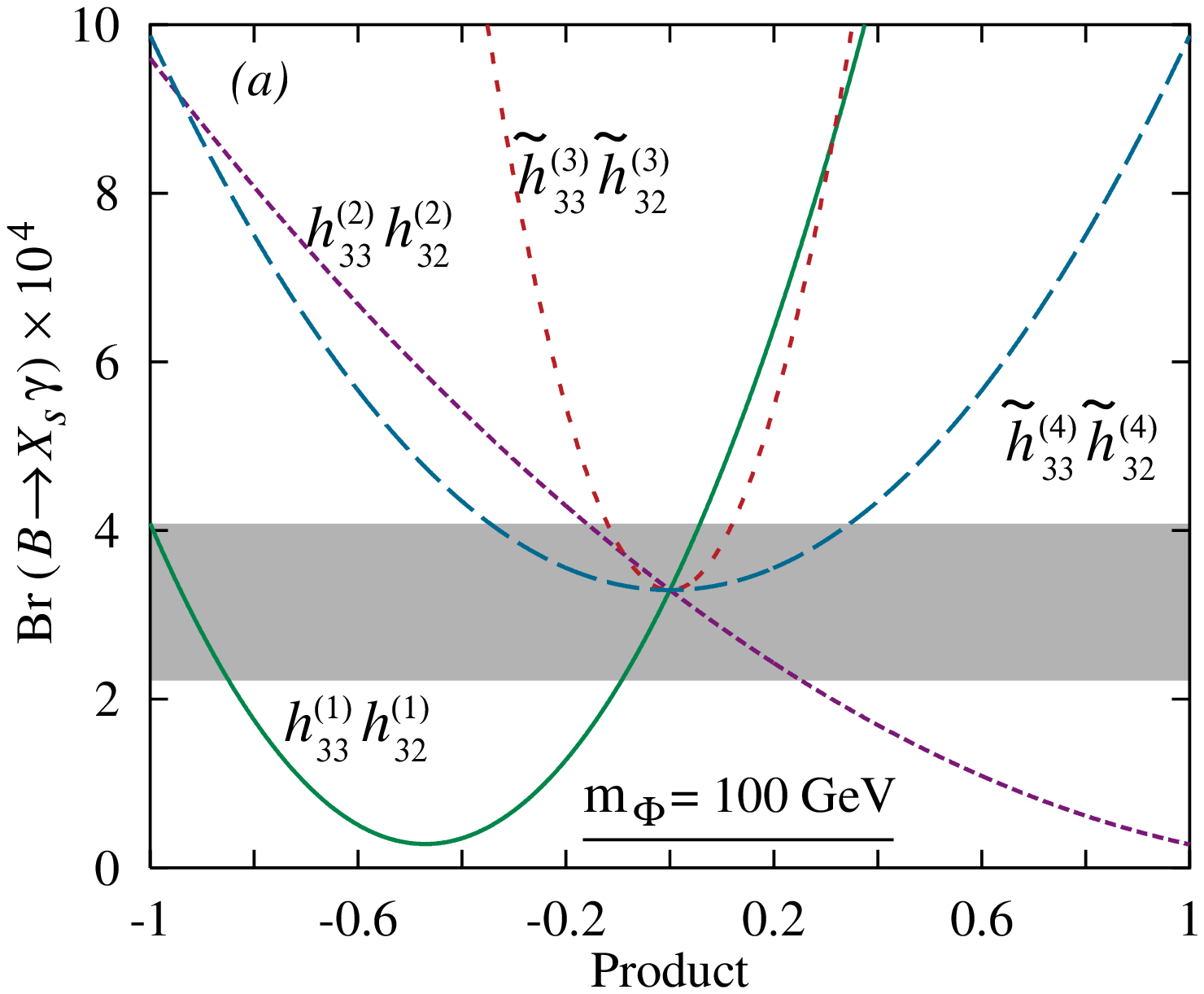}
	\hspace*{-2ex}
\epsfxsize=6.5cm\epsfysize=6.0cm
                     \epsfbox{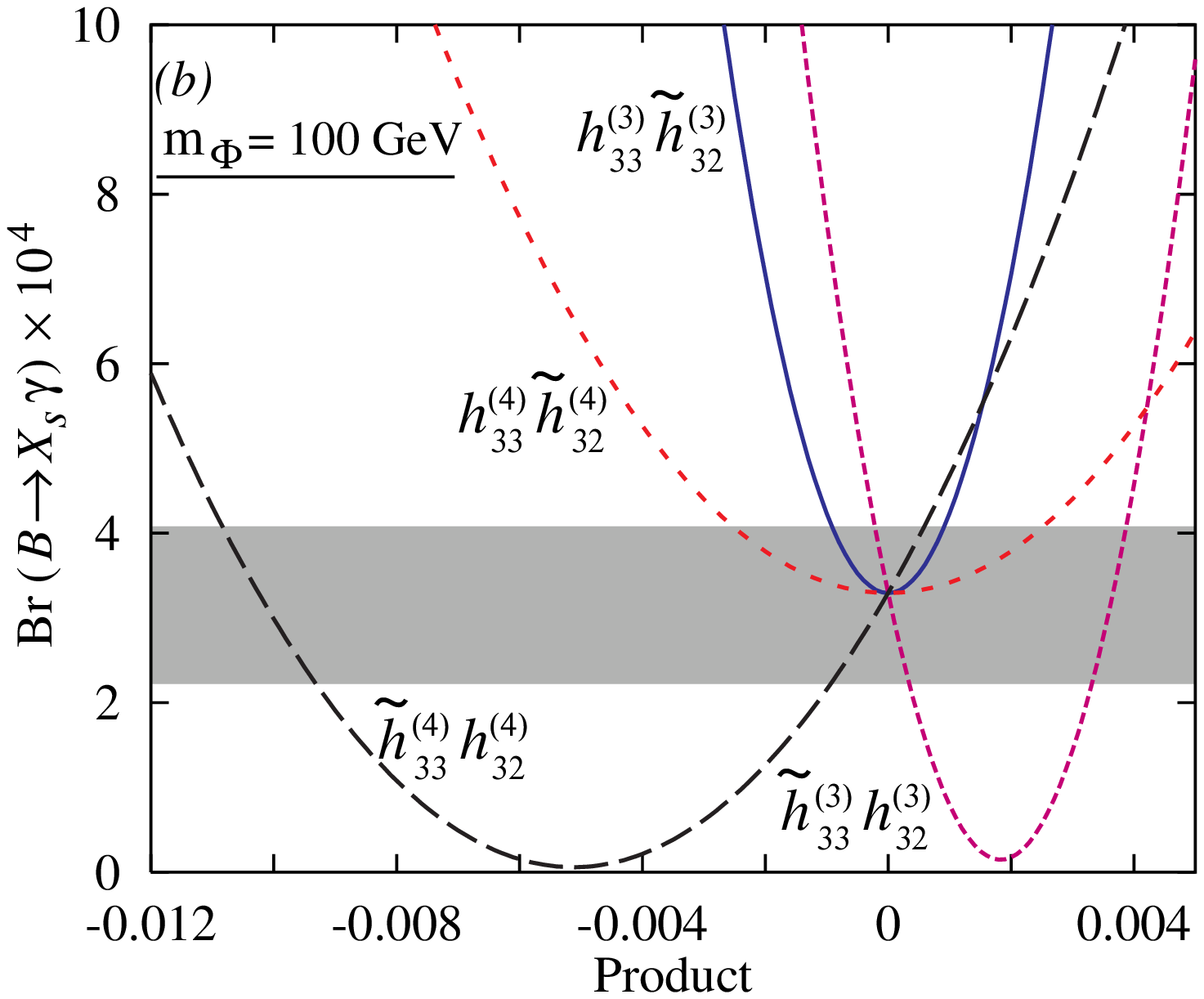}
}
  \caption{\em The partial width for $B \to X_s \gamma$ 
	       as a function of the product of the diquark and/or 
                R parity violating
	       couplings for a fixed diquark mass of 100 GeV,
	       The shaded region represents the $1 \sigma$ limits 
	       of the experimentally observed value.
	       The curves for $h^{(3)}_{33} h^{(3)}_{32}$
		and $h^{(4)}_{33} h^{(4)}_{32}$ are identical 
		to those for $h^{(1)}_{33} h^{(1)}_{32}$
	       and $h^{(2)}_{33} h^{(2)}_{32}$ respectively.
	  }
\label{fig:hhvar}
\end{figure}
%%%%%%%%%%%%%%%%%%
In the absence of an $L$-violating coupling, these diquarks clearly do not
influence the semileptonic decay modes of the $B$-meson. Thus, we 
may continue to normalize the radiative $b$-decay against 
$b \to c e {\bar \nu_e}$ in order to avoid the severe 
dependence on $m_b$. 
In Fig.~\ref{fig:hhvar}, we plot the branching ratio $BR(B \to X_s \gamma)$
in presence of a diquark (multiplet) of mass $100 \gev$. 
We continue to work under the assumption that only one pair of couplings 
are non-zero. Furthermore, we assume the said couplings to be 
real\footnote{The extension to complex couplings is straightforward. 
	      The imaginary parts, however, can be better constrained 
	      from an analysis of the $CP$ violating 
	      decay modes.}.
That the curves should be parabolic in the product of the two couplings 
in question is obvious. 
To appreciate the fact that many of these curves 
have their minima lying on the SM value, one needs to consider the 
chirality structure of the corresponding diquark couplings 
(see Table~\ref{tab:consts}).
For example, combinations involving either of 
$\tilde h^{(3)}_{32}$ or $\tilde h^{(4)}_{32}$, imply 
that there is no left-handed coupling 
to the strange quark. From eqs.(\ref{Wilsons}), it is then easy to 
see that the SM Wilson coefficients $C_{7,8}$ remain unaffected. 
Consequently, the new contribution adds {\em incoherently} with the 
SM amplitude. For the rest of the combinations, though, the 
interference term is non-negligible leading to a shift in the 
minimum. Hence, unlike those for the first set of combinations 
(those involving $\tilde h^{(3,4)}_{32}$), the branching 
ratios corresponding 
to these sets are in agreement with the experimental numbers
for {\em two non-contiguous} ranges of the product. 

In each of  eqs.(\ref{Wilsons}), the second term, whenever allowed, 
 is clearly the dominant piece.
 This enhancement by the factor of $m_t/ m_b$ 
comes into play only when the diquark 
couplings to the bottom and the strange quarks have pieces with 
opposite chirality. In other words, for a diquark of the type 
$\Phi_3$ (or $\Phi_4$), the simultaneous presence of both the
allowed types of couplings is severely constrained by the data
on $B \to X_s \gamma$.

%%%%%
\input{table.limits}
%%%%%
In Table~\ref{tab:limits}, we capture the essence of 
Fig.~\ref{fig:hhvar} in the form of actual limits 
that can be set on such products of couplings, for a 
diquark mass of $100 \gev$.
Understandably, the $2 \sigma$ bounds are weaker 
than the $1 \sigma$ ones. 
Similarly, the color-sextet couplings are more severely 
constrained than the color-triplet ones.
It should be noted that 
the structure of the interaction terms $h^{(1)}_{ij}$ are 
the same as those for $h^{(3)}_{ij}$.
 Consequently, the bounds are exactly the same. 
A similar story obtains for $h^{(2)}_{ij}$ and $h^{(4)}_{ij}$. 
As discussed above, for a few of the products there are two
non-contiguous bands allowed. 
For the combinations $h_{33}^{(2)}h_{32}^{(2)}$ and 
$h_{33}^{(4)}h_{32}^{(4)}$, though, 
the second window (both at the $1 \sigma$ and $2 \sigma$ levels)
lies beyond the perturbative limit and, hence,  are 
phenomenologically uninteresting.

As discussed earlier, ${\tilde h}_{ij}^{(4)}$ is
analogous to the trilinear $R$-parity violating coupling 
$\lambda^{\prime \prime}_{ijk}$.
Thus the constraints on ${\tilde h}_{33}^{(4)} {\tilde h}_{32}^{(4)}$ 
are equivalent
to those on the product 
$\lambda^{\prime \prime}_{3j2} \lambda^{\prime \prime}_{3j3}$. 
For each of these couplings, the best {\em individual} 
bound comes from the precision measurements at the $Z$ pole~\cite{b_c_s,ggw},
and amounts to 
$\lambda^{\prime \prime}_{3j2} \ , \lambda^{\prime \prime}_{3j3} < 0.50$
at the $1 \sigma$ level. We thus do not do very well as far as this 
particular combination is concerned. This can be attributed to both the 
chirality and the color structure of the operator, each of which is
``unfavourable'' as far the $b \to s \gamma$ decay is concerned. 
For most of the other combinations though, we do {\em significantly better} 
than the product of individual bounds~\cite{b_c_s}.

%%%%%%%%%%%%%%%%%%%%%%%%%%%%%%%%%%%%%%

\begin{figure}[h]
\centerline{
\epsfxsize=6.5cm\epsfysize=6.0cm
                     \epsfbox{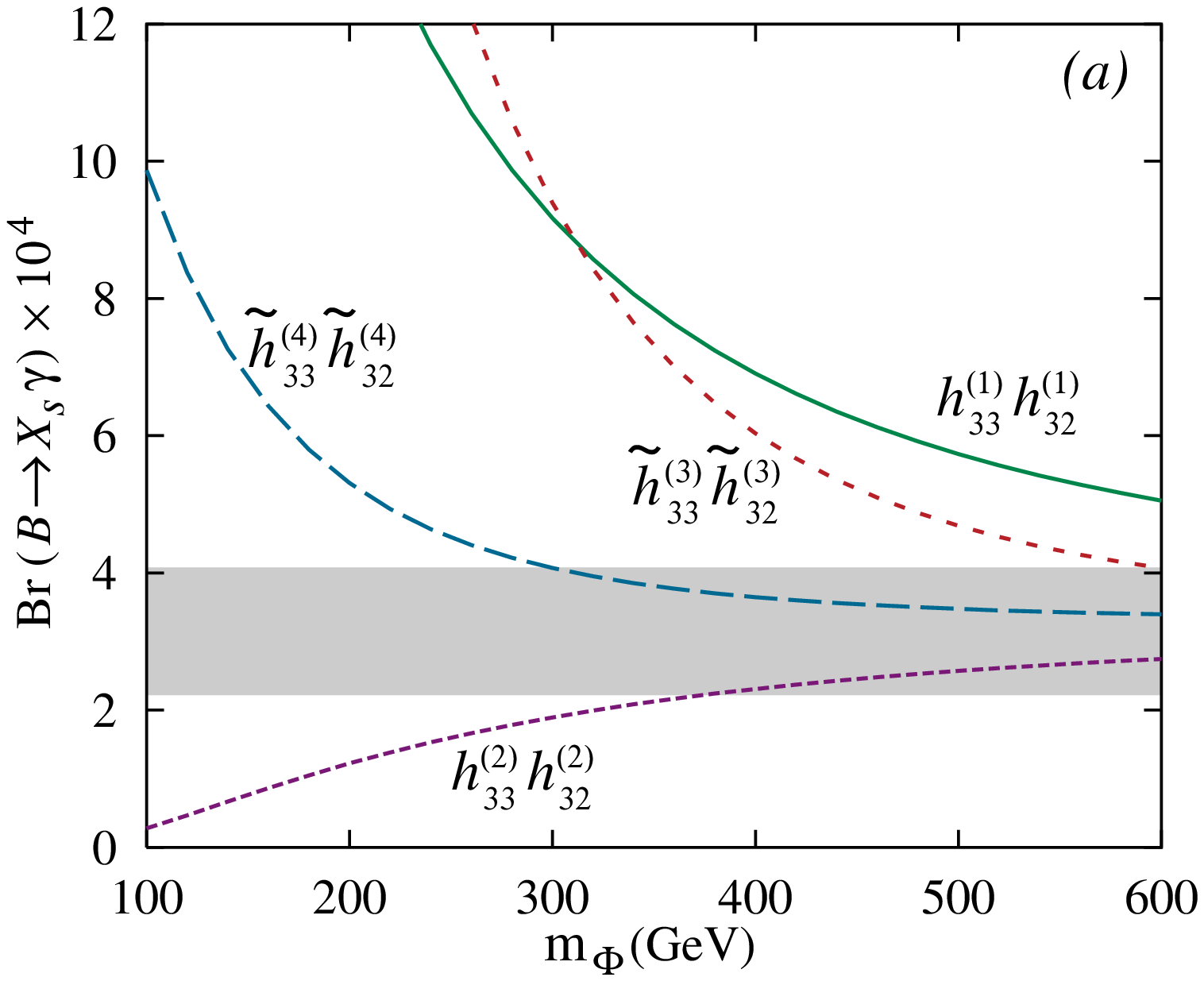}
	\hspace*{-2ex}
\epsfxsize=6.5cm\epsfysize=6.0cm
                     \epsfbox{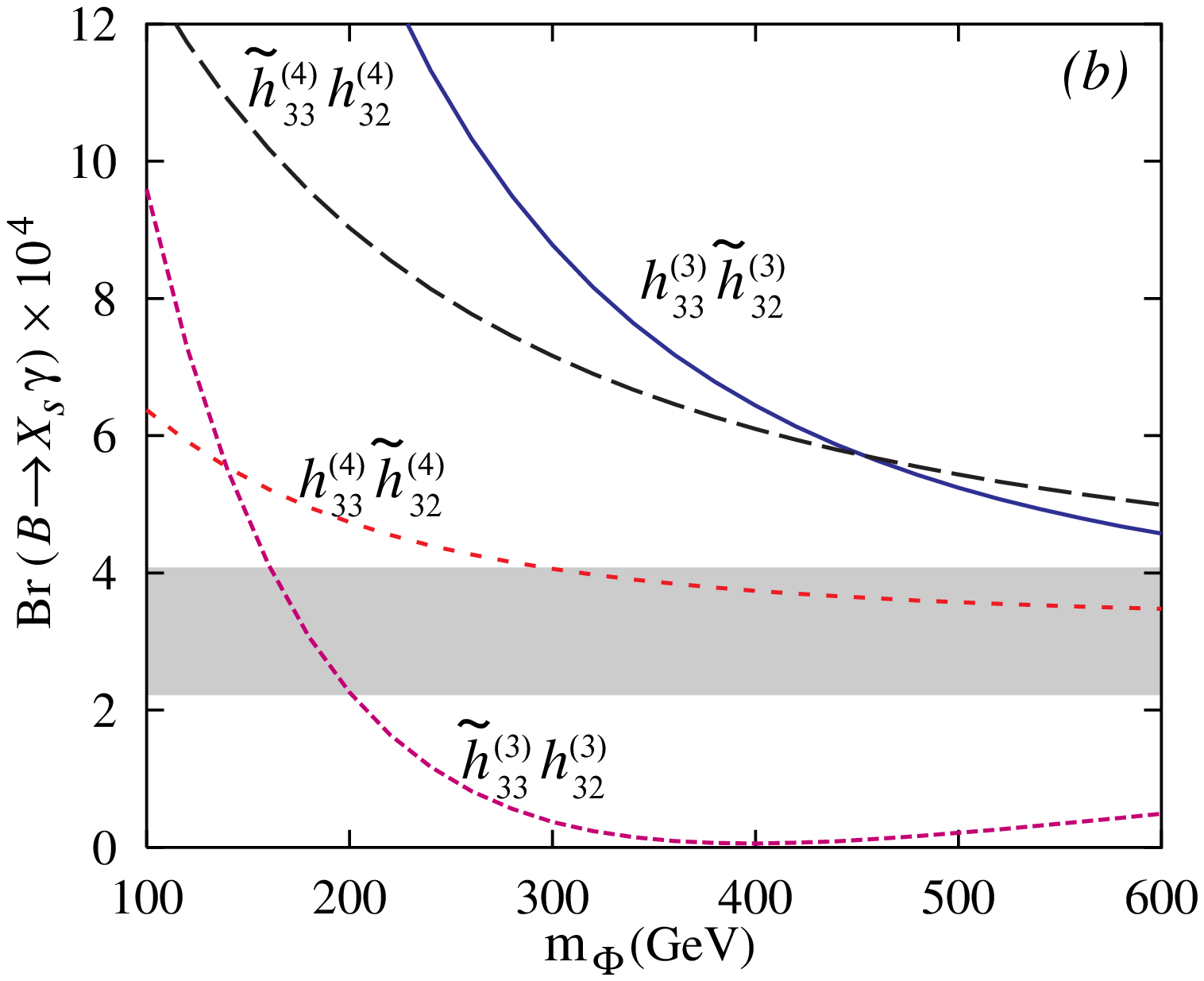}
}
  \caption{\em The partial width for $B \to X_s \gamma$ 
	       as a function of diquark mass. 
	       {\em (a)} For each curve, 
	       the associated product of diquark and/or R parity violating 
                couplings 
	       is held to be 1, while all other couplings 
	       are set to be vanishingly small. The shaded 
	       region represents the $1 \sigma$ limits 
	       of the observed value.
	       The curves for $h^{(3)}_{33} h^{(3)}_{32}$
		and $h^{(4)}_{33} h^{(4)}_{32}$ are identical 
		to those for $h^{(1)}_{33} h^{(1)}_{32}$
	       and $h^{(2)}_{33} h^{(2)}_{32}$ respectively.
	       {\em (b)} As in {\em (a)}, but 
	       the non-zero products of diquark and/or R parity 
               voilating  couplings 
	       are held to be 0.005.
	  }
\label{fig:mvar}
\end{figure}

%%%%%%%%%%%%%%%%%%%%%%%%%%%%%%%%%%5

In our effort to compare with the results available in the 
literature,  we have, until now, held the diquark mass to be 
$100 \gev$ and varied the strength of its coupling. In reality,
though a diquark is more likely to be somewhat heavier. 
For the sake of completeness, we next investigate the dependence 
on the diquark mass (Figs.~\ref{fig:mvar}), while holding the 
product fixed. As is expected, the extra contribution falls off 
with $m_\Phi$. The fall-off is somewhat slower than $m_{\Phi}^{-2}$ 
(see the expressions for $F_i(y)$ in eq.(\ref{the_f's})) and 
the effects persist till $\sim 3~ TeV$. The different rates of fall-off
are governed by the dominant $F_i(y)$ in each case. 
The case for the combination ${\tilde h}_{33}^{(3)}h_{32}^{(3)}$ 
 looks somewhat nontrivial. However, the shape is just a 
consequence of accidental cancellations between various 
terms of eq.(\ref{bs1}). As the exact nature of these cancellations depend 
crucially on the value of the diquark couplings, not much should 
be read into the shape in general or the minimum in particular.

\begin{figure}[hb]
\centerline{
\epsfxsize=6.5cm\epsfysize=6.0cm
                     \epsfbox{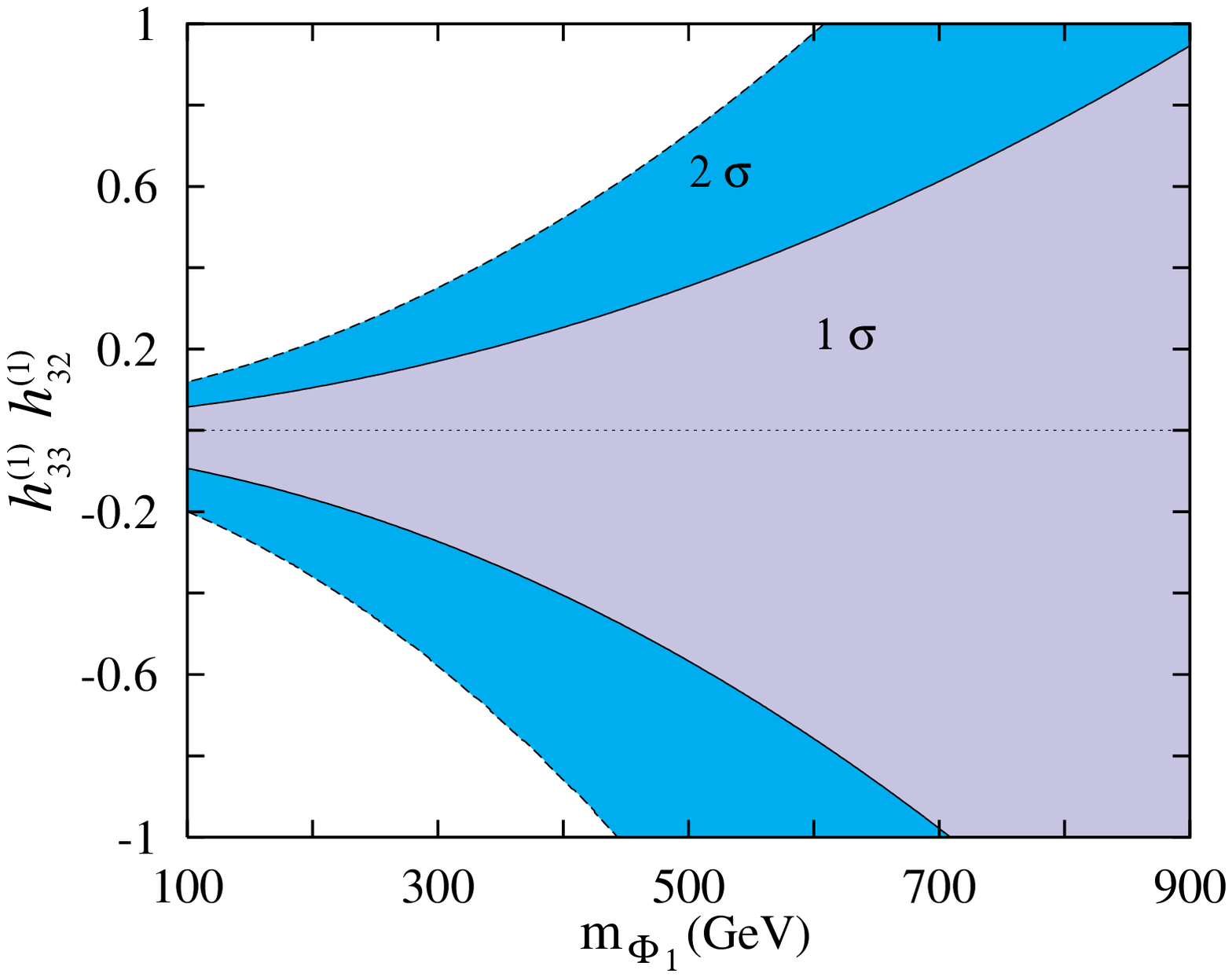}
	\hspace*{-2ex}
\epsfxsize=6.5cm\epsfysize=6.0cm
                     \epsfbox{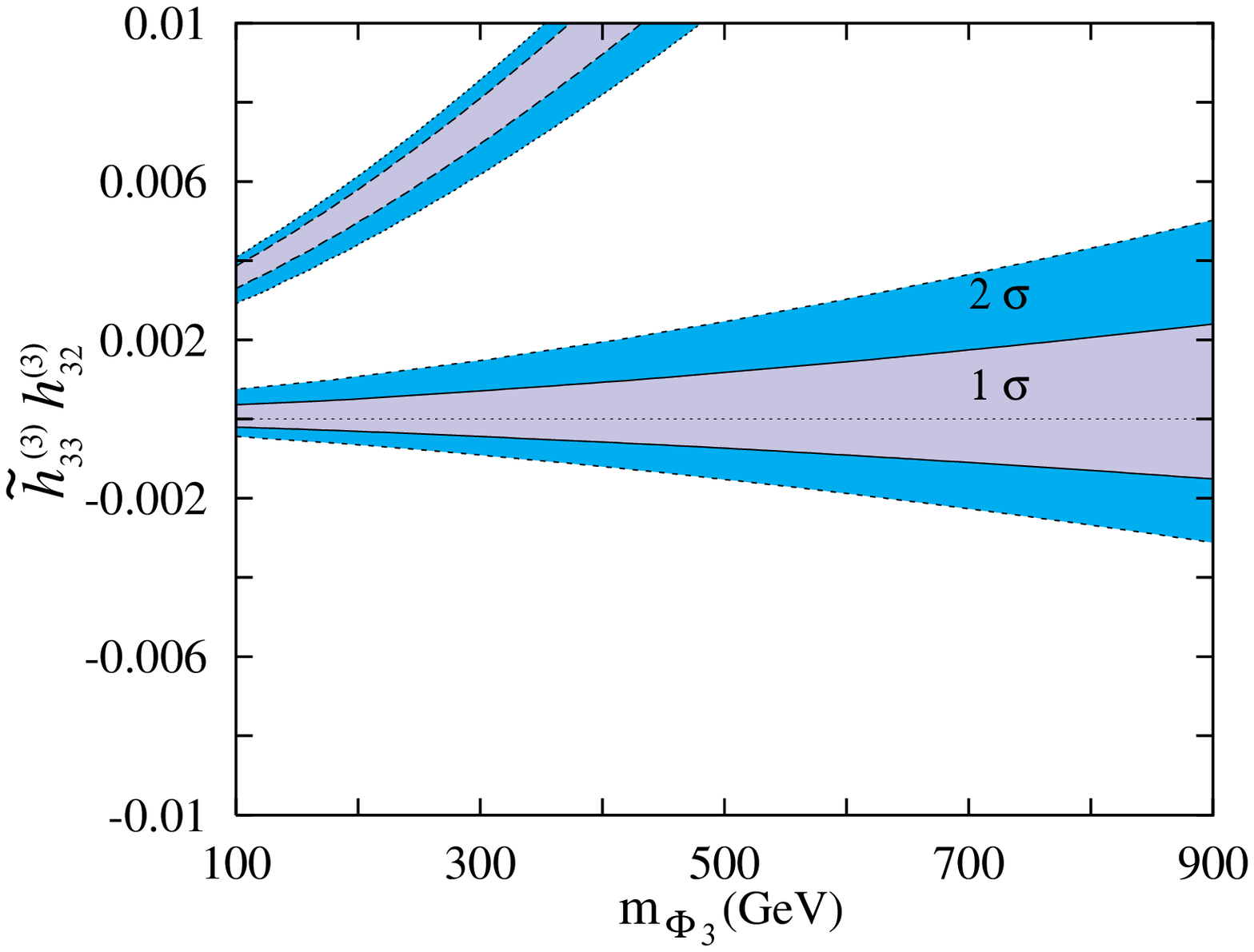}
}
\caption{\em The region of the parameter space allowed by the data 
	when all other couplings are set to zero. The lightly shaded 
	area agrees with the data at $1 \sigma$ level, whereas 
	the encompassing darker region agrees at $2 \sigma$.
	}
\label{fig:cont}
\end{figure}

The two dependences ($m_\Phi$ and coupling strength) that we have 
studied can be combined to rule out parts of the phase space. In 
Figs.\ref{fig:cont}, we exhibit this for two particular combinations. 
In each, the shaded regions of the parameter space are in agreement 
with the experimental results at the designated level. 
For $h_{33}^{(1)}h_{32}^{(1)}$, the second allowed region is
beyond the perturbative limit.

In summary, we have studied the effects of the scalar diquark 
and/or R prity violating coupling
to the 
branching ratio $B \to X_s \gamma$. Among the possible new contributions, 
the scalar diquark mediated
diagram yield promising effects. The precise measurement of this 
branching ratio
at the upcoming B factories in near future and the reduction of 
theoretical uncertainty will improve the limits on the product combination 
of different scalar diquark and/or R parity violating couplings, we obtained.

\vskip 25pt
\begin{center}
{\bf Acknowledgement }
\end{center}
D. Choudhury acknowledges the Department of Science and Technology, India 
for the Swarnajayanti Fellowship grant. 

\vspace*{.2 in}

%%%%%%%%%%%%%%%%%%%%%%%%%%%%%%%%%%%%%%%%%%%%%%%%%%%%%%%%%%%%%%%%%%%%%%%%
%								       %
%    Journal macros	(in the Nucl Phys style)                       %
%								       %
%    To change over to Phys. Rev style, replace 		       %
%		(#3) #2       by    , #2 (#3)			       %
%								       %
\newcommand{\ib}[3]   {{\em ibid.\/} {\bf #1} (#3) #2}		       %
\newcommand{\app}[3]  {{\em Acta Phys. Polon.	B\/}{\bf #1} (#3) #2}  %
\newcommand{\ajp}[3]  {{\em Am. J. Phys.\/} {\bf #1} (#3) #2}	       %
\newcommand{\ap}[3]   {{\em Ann. Phys.	(NY)\/}	{\bf #1} (#3)	#2}    %
\newcommand{\araa}[3] {{\em Annu. Rev. Astron. Astrophys.\/}	       %
          {\bf#1} (#3) #2}					       %
\newcommand{\apj}[3]  {{\em Astrophys. J.\/} {\bf #1} (#3) #2}         %
\newcommand{\apjs}[3] {{\em Astrophys. J. Suppl.\/}                    %
          {\bf	#1} (#3) #2}			                       %
\newcommand{\apjl}[3] {{\em Astrophys. J. Lett.\/} {\bf #1} (#3) #2}   %
\newcommand{\astropp}[3]{Astropart. Phys. {\bf #1} (#3) #2}	       %
\newcommand{\eur}[3]  {Eur. Phys. J. {\bf C#1} (#3) #2}                %
\newcommand{\iauc}[4] {{\em IAU Circular\/} #1                         %
       (\ifcase#2\or January \or February \or March  \or April \or May %
                 \or June    \or July     \or August \or September     %
                 \or October \or November \or December                 %
        \fi \ #3, #4)}					               %
\newcommand{\ijmp}[3] {Int. J. Mod. Phys. {\bf A#1} (#3) #2}           %
\newcommand{\jetp}[6] {{\em Zh. Eksp. Teor. Fiz.\/} {\bf #1} (#3) #2   %
     [English translation: {\it Sov. Phys.--JETP } {\bf #4} (#6) #5]}  %
\newcommand{\jetpl}[6]{{\em ZhETF Pis'ma\/} {\bf #1} (#3) #2           %
     [English translation: {\it JETP Lett.\/} {\bf #4} (#6) #5]}       %
\newcommand{\jhep}[3] {JHEP {\bf #1} (#3) #2}                          %
\newcommand{\mpla}[3] {Mod. Phys. Lett. {\bf A#1} (#3) #2}             %
\newcommand{\nat}[3]  {Nature (London) {\bf #1} (#3) #2}	       %
\newcommand{\nuovocim}[3]{Nuovo Cim. {\bf #1} (#3) #2}	               %
\newcommand{\np}[3]   {Nucl. Phys. {\bf B#1} (#3) #2}		       %
\newcommand{\npbps}[3]{Nucl. Phys. B (Proc. Suppl.)                    %
           {\bf #1} (#3) #2}	                                       %
\newcommand{\philt}[3] {Phil. Trans. Roy. Soc. London A {\bf #1} #2    %
	(#3)}							       %
\newcommand{\prev}[3] {Phys. Rev. {\bf #1} (#3) #2}	       	       %
\newcommand{\prd}[3]  {{Phys. Rev.}{\bf D#1} (#3) #2}		       %
\newcommand{\prl}[3]  {Phys. Rev. Lett. {\bf #1} (#3) #2}	       %
\newcommand{\plb}[3]  {{Phys. Lett.} {\bf B#1} (#3) #2}		       %
\newcommand{\prep}[3] {Phys. Rep. {\bf #1} (#3) #2}		       %
\newcommand{\ptp}[3]  {Prog. Theoret. Phys. (Kyoto) {\bf #1} (#3) #2}  %
\newcommand{\rpp}[3]  {Rep. Prog. Phys. {\bf #1} (#3) #2}              %
\newcommand{\rmp}[3]  {Rev. Mod. Phys. {\bf #1} (#3) #2}               %
\newcommand{\sci}[3]  {Science {\bf #1} (#3) #2}		       %
\newcommand{\zp}[3]   {Z.~Phys. C{\bf#1} (#3) #2}		       %
\newcommand{\uspekhi}[6]{{\em Usp. Fiz. Nauk.\/} {\bf #1} (#3) #2      %
     [English translation: {\it Sov. Phys. Usp.\/} {\bf #4} (#6) #5]}  %
\newcommand{\yadfiz}[4]{Yad. Fiz. {\bf #1} (#3) #2 [English	       %
	transl.: Sov. J. Nucl.	Phys. {\bf #1} #3 (#4)]}	       %
\newcommand{\hepph}[1] {(electronic archive:	hep--ph/#1)}	       %
\newcommand{\hepex}[1] {(electronic archive:	hep--ex/#1)}	       %
\newcommand{\astro}[1] {(electronic archive:	astro--ph/#1)}	       %
%	\relax							       %
%	%%%	End	Journal	macro definitions		       %
%								       %
%%%%%%%%%%%%%%%%%%%%%%%%%%%%%%%%%%%%%%%%%%%%%%%%%%%%%%%%%%%%%%%%%%%%%%%%

\end{document}

%% file: table.qnos
\begin{table}[h]
\begin{center}
\begin{tabular}{|c|c|c|}
\hline
& &  \\[-2ex]
 Diquark Type  &  Coupling & $SU(3)_c\times SU(2)_L\times U(1)_Y$\\
& &  \\[-2ex]
 \hline
 \hline
& &  \\[-2ex]
 $\Phi_1$ & $h_{ij}^{(1)} ({\bar Q}_{L i})^c Q_{L j} \Phi_1$ & 
	($ {\bar 6}$, $3$, $-{2\over 3}$)\\ 
& &  \\[-2ex]
 \hline
& &  \\[-2ex]
 $\Phi_2$ & $h_{ij}^{(2)} ({\bar Q}_{L i})^c Q_{L j} \Phi_2$ & 
	($3$, $3$, $-{2\over 3}$)\\ 
& &  \\[-2ex]
\hline
& &  \\[-2ex]
 $\Phi_3$ & 
   $\left[ h_{ij}^{(3)} ({\bar Q}_{L i})^c Q_{L j} 
         + \tilde h_{ij}^{(3)}({\bar u}_{R i})^c d_{R j}
    \right] \: \Phi_3 $ & ($ {\bar 6}$, $1$, $-{2\over 3}$)\\ 
& &  \\[-2ex]
\hline
& &  \\[-2ex]
 $\Phi_4$ & 
    $\left[ h_{ij}^{(4)} ({\bar Q}_{L i})^c Q_{L j} 
          + \tilde h_{ij}^{(4)}({\bar u}_{R i})^c d_{R j}
     \right] \Phi_4 $ & ($ 3$, $1$, $-{2\over 3}$)\\
& &  \\[-2ex]
 \hline
& &  \\[-2ex]
 $\Phi_5$ & $h_{ij}^{(5)} ({\bar u}_{R i})^c u_{R j}
 \Phi_5$ & ($ {\bar 6}$, $1$, $-{8\over 3}$)\\
& &  \\[-2ex]
 \hline
& &  \\[-2ex]
 $\Phi_6$ & $h_{ij}^{(6)} ({\bar u}_{R i})^c u_{R j}
 \Phi_6$ & ($ 3$, $1$, $-{8\over 3}$)\\
& &  \\[-2ex]
 \hline
& &  \\[-2ex]
 $\Phi_7$ & $h_{ij}^{(7)} ({\bar d}_{R i})^c d_{R j}
 \Phi_7$ & ($ \bar 6$, $1$, ${4\over 3}$)\\
& &  \\[-2ex]
 \hline
& &  \\[-2ex]
 $\Phi_8$ & $h_{ij}^{(8)} ({\bar d}_{R i})^c d_{R j}
 \Phi_8$ & ($3$, $1$, ${4\over 3}$)
	\\
 \hline
\end{tabular}
\end{center}
    \caption{\em Gauge Quantum Numbers and Yukawa Couplings 
	         of Scalar Diquarks $(Q_{em} = T_3 +{Y \over 2})$.
            }
	\label{tab:qnos}
\end{table}

%% file: fig_bsphot
%%%%%%%%%%%%%%%%%%%%%%%%1%%%%%%%%%%%%%%%%%%%%%%%%%%%%%%%%%%%%%%%%
\begin{figure}[h]
\begin{center}
\begin{picture}(350,205)(0,5)
\ArrowLine(0,200)(10,200)
\ArrowLine(10,200)(20,200)
\Text(5,204)[rb]{$\footnotesize{b}$}
\ArrowLine(60 ,200)(80,200)
\Photon(10,200)(10,150){2}{8}
\Text(0,160)[lb]{$\footnotesize{\gamma}$}
\Text(75,204)[lb]{$\footnotesize{s}$}
\Photon(20,200)(60,200){2}{8}
\Text(40,207)[cb]{$\footnotesize{W}$}
\ArrowArc(40,200)(20,180,0)
\Text(40,172)[ct]{$\footnotesize{t}$}
\Text(40,150)[ct]{$\footnotesize{(a)}$}
\ArrowLine(90,200)(110,200)
\Text(95,204)[rb]{$\footnotesize{b}$}
\ArrowLine(150,200)(160,200)
\ArrowLine(160,200)(170,200)
\Photon(160,200)(160,150){2}{8}
\Text(150,160)[lb]{$\footnotesize{\gamma}$}
\Text(165,204)[lb]{$\footnotesize{s}$}
\Photon(110,200)(150,200){2}{8}
\Text(130,207)[cb]{$\footnotesize{W}$}
\ArrowArc(130,200)(20,180,0)
\Text(130,172)[ct]{$\footnotesize{t}$}
\Text(130,150)[ct]{$\footnotesize{(b)}$}
\ArrowLine(180,200)(200,200)
\Text(185,204)[rb]{$\footnotesize{b}$}
\ArrowLine(240,200)(260,200)
\Photon(220,180)(220,150){2}{4}
\Text(210,160)[lb]{$\footnotesize{\gamma}$}
\Text(255,204)[lb]{$\footnotesize{s}$}
\Photon(200,200)(240,200){2}{8}
\Text(220,207)[cb]{$\footnotesize{W}$}
\ArrowArc(220,200)(20,180,270)
\ArrowArc(220,200)(20,270,0)
\Text(200,185)[ct]{$\footnotesize{t}$}
\Text(240,185)[ct]{$\footnotesize{t}$}
\Text(220,150)[ct]{$\footnotesize{(c)}$}
\ArrowLine(270,200)(290,200)
\Text(275,204)[rb]{$\footnotesize{b}$}
\ArrowLine(290,200)(330,200)
\ArrowLine(330,200)(350,200)
\Photon(310,180)(310,150){2}{4}
\Text(300,160)[lb]{$\footnotesize{\gamma}$}
\Text(325,204)[lb]{$\footnotesize{s}$}
\Text(310,204)[cb]{$\footnotesize{t}$}
\PhotonArc(310,200)(20,180,270){2}{6}
\PhotonArc(310,200)(20,270,0){2}{6}
\Text(295,180)[ct]{$\footnotesize{W}$}
\Text(330,180)[ct]{$\footnotesize{W}$}
\Text(310,150)[ct]{$\footnotesize{(d)}$}
\ArrowLine(0,100)(10,100)
\ArrowLine(10,100)(20,100)
\Text(5,104)[rb]{$\footnotesize{b}$}
\ArrowLine(60 ,100)(80,100)
\Photon(10,100)(10,50){2}{8}
\Text(0,60)[lb]{$\footnotesize{\gamma}$}
\Text(75,104)[lb]{$\footnotesize{s}$}
\DashArrowLine(60,100)(20,100){2}
\Text(40,107)[cb]{$\footnotesize{\Phi_i}$}
\ArrowArcn(40,100)(20,0,180)
\Text(40,72)[ct]{$\footnotesize{t}$}
\Text(40,50)[ct]{$\footnotesize{(e)}$}
\ArrowLine(90,100)(110,100)
\Text(95,104)[rb]{$\footnotesize{b}$}
\ArrowLine(150,100)(160,100)
\ArrowLine(160,100)(170,100)
\Photon(160,100)(160,50){2}{8}
\Text(150,60)[lb]{$\footnotesize{\gamma}$}
\Text(165,104)[lb]{$\footnotesize{s}$}
\DashArrowLine(150,100)(110,100){2}
\Text(130,107)[cb]{$\footnotesize{\Phi_i}$}
\ArrowArcn(130,100)(20,0,180)
\Text(130,72)[ct]{$\footnotesize{t}$}
\Text(130,50)[ct]{$\footnotesize{(f)}$}
\ArrowLine(180,100)(200,100)
\Text(185,104)[rb]{$\footnotesize{b}$}
\ArrowLine(240,100)(260,100)
\Photon(220,80)(220,50){2}{4}
\Text(210,60)[lb]{$\footnotesize{\gamma}$}
\Text(255,104)[lb]{$\footnotesize{s}$}
\DashArrowLine(240,100)(200,100){2}
\Text(220,107)[cb]{$\footnotesize{\Phi_i}$}
\ArrowArcn(220,100)(20,270,180)
\ArrowArcn(220,100)(20,0,270)
\Text(200,85)[ct]{$\footnotesize{t}$}
\Text(240,85)[ct]{$\footnotesize{t}$}
\Text(220,50)[ct]{$\footnotesize{(g)}$}
\ArrowLine(270,100)(290,100)
\Text(275,104)[rb]{$\footnotesize{b}$}
\ArrowLine(330,100)(290,100)
\ArrowLine(330,100)(350,100)
\Photon(310,80)(310,50){2}{4}
\Text(300,60)[lb]{$\footnotesize{\gamma}$}
\Text(325,104)[lb]{$\footnotesize{s}$}
\Text(310,104)[cb]{$\footnotesize{t}$}
\DashArrowArcn(310,100)(20,270,180){2}
\DashArrowArcn(310,100)(20,0,270){2}
\Text(295,80)[ct]{$\footnotesize{\Phi_i}$}
\Text(330,80)[ct]{$\footnotesize{\Phi_i}$}
\Text(310,50)[ct]{$\footnotesize{(h)}$}
\end{picture} 
\end{center}
\vspace*{-7ex}
      \caption{\em Feynman diagrams that determine the one loop 
		   $b \to s \gamma$ decay amplitude.}
	\label{fig:feyn_bsphot}
\end{figure}

%% file: fig_bsgl
\begin{figure}[h]
\begin{center}
\begin{picture}(350,205)(-5,5)
\ArrowLine(0,200)(10,200)
\ArrowLine(10,200)(20,200)
\Text(5,204)[rb]{$\footnotesize{b}$}
\ArrowLine(60 ,200)(80,200)
\Gluon(10,200)(10,150){2}{8}
\Text(-5,160)[lb]{$\footnotesize{G}$}
\Text(75,204)[lb]{$\footnotesize{s}$}
\Photon(20,200)(60,200){2}{8}
\Text(40,207)[cb]{$\footnotesize{W}$}
\ArrowArc(40,200)(20,180,0)
\Text(40,172)[ct]{$\footnotesize{t}$}
\Text(40,150)[ct]{$\footnotesize{(a)}$}
\ArrowLine(90,200)(110,200)
\Text(95,204)[rb]{$\footnotesize{b}$}
\ArrowLine(150,200)(160,200)
\ArrowLine(160,200)(170,200)
\Gluon(160,200)(160,150){2}{8}
\Text(145,160)[lb]{$\footnotesize{G}$}
\Text(165,204)[lb]{$\footnotesize{s}$}
\Photon(110,200)(150,200){2}{8}
\Text(130,207)[cb]{$\footnotesize{W}$}
\ArrowArc(130,200)(20,180,0)
\Text(130,172)[ct]{$\footnotesize{t}$}
\Text(130,150)[ct]{$\footnotesize{(b)}$}
\ArrowLine(180,200)(200,200)
\Text(185,204)[rb]{$\footnotesize{b}$}
\ArrowLine(240,200)(260,200)
\Gluon(220,180)(220,150){2}{4}
\Text(205,160)[lb]{$\footnotesize{G}$}
\Text(255,204)[lb]{$\footnotesize{s}$}
\Photon(200,200)(240,200){2}{8}
\Text(220,207)[cb]{$\footnotesize{W}$}
\ArrowArc(220,200)(20,180,270)
\ArrowArc(220,200)(20,270,0)
\Text(200,185)[ct]{$\footnotesize{t}$}
\Text(240,185)[ct]{$\footnotesize{t}$}
\Text(220,150)[ct]{$\footnotesize{(c)}$}
\ArrowLine(0,100)(10,100)
\ArrowLine(10,100)(20,100)
\Text(5,104)[rb]{$\footnotesize{b}$}
\ArrowLine(60 ,100)(80,100)
\Gluon(10,100)(10,50){2}{8}
\Text(-5,60)[lb]{$\footnotesize{G}$}
\Text(75,104)[lb]{$\footnotesize{s}$}
\DashArrowLine(60,100)(20,100){2}
\Text(40,107)[cb]{$\footnotesize{\Phi_i}$}
\ArrowArcn(40,100)(20,0,180)
\Text(40,72)[ct]{$\footnotesize{t}$}
\Text(40,50)[ct]{$\footnotesize{(d)}$}
\ArrowLine(90,100)(110,100)
\Text(95,104)[rb]{$\footnotesize{b}$}
\ArrowLine(150,100)(160,100)
\ArrowLine(160,100)(170,100)
\Gluon(160,100)(160,50){2}{8}
\Text(145,60)[lb]{$\footnotesize{G}$}
\Text(165,104)[lb]{$\footnotesize{s}$}
\DashArrowLine(150,100)(110,100){2}
\Text(130,107)[cb]{$\footnotesize{\Phi_i}$}
\ArrowArcn(130,100)(20,0,180)
\Text(130,72)[ct]{$\footnotesize{t}$}
\Text(130,50)[ct]{$\footnotesize{(e)}$}
\ArrowLine(180,100)(200,100)
\Text(185,104)[rb]{$\footnotesize{b}$}
\ArrowLine(240,100)(260,100)
\Gluon(220,80)(220,50){2}{4}
\Text(205,60)[lb]{$\footnotesize{G}$}
\Text(255,104)[lb]{$\footnotesize{s}$}
\DashArrowLine(240,100)(200,100){2}
\Text(220,107)[cb]{$\footnotesize{\Phi_i}$}
\ArrowArcn(220,100)(20,270,180)
\ArrowArcn(220,100)(20,0,270)
\Text(200,85)[ct]{$\footnotesize{t}$}
\Text(240,85)[ct]{$\footnotesize{t}$}
\Text(220,50)[ct]{$\footnotesize{(f)}$}
\ArrowLine(270,100)(290,100)
\Text(275,104)[rb]{$\footnotesize{b}$}
\ArrowLine(330,100)(290,100)
\ArrowLine(330,100)(350,100)
\Gluon(310,80)(310,50){2}{4}
\Text(295,60)[lb]{$\footnotesize{G}$}
\Text(325,104)[lb]{$\footnotesize{s}$}
\Text(310,104)[cb]{$\footnotesize{t}$}
\DashArrowArcn(310,100)(20,270,180){2}
\DashArrowArcn(310,100)(20,0,270){2}
\Text(295,80)[ct]{$\footnotesize{\Phi_i}$}
\Text(330,80)[ct]{$\footnotesize{\Phi_i}$}
\Text(310,50)[ct]{$\footnotesize{(g)}$}
\end{picture} \\
\end{center}
\vspace*{-7ex}
      \caption{\em Feynman diagrams that determine the one loop 
	           $b \to s G$ decay amplitude.}
	\label{fig:feyn_bsglue}
\end{figure}

%% file: table.consts
\begin{table}[ht]
\begin{center}
\begin{tabular}{|c|c|c|c|c|c|c|}\hline
& & & & & &  \\[-2ex]
 Diquark Type  &  $C_{\Phi}$ & $C_t$ & $l_b$ & $l_s$ & $r_b$ & $r_s$\\
& & & & & &  \\[-2ex]
 \hline
 \hline
& & & & & &  \\[-2ex]
% & & & &  \\ \hline
 $\Phi_1$ &$ -5/2$ & $1/2$  & 
	$h_{33}^{(1)}/\sqrt{2}$  & $h_{32}^{(1)}/\sqrt{2}$ & 0 & 0
	\\
& & & & & &  \\[-2ex]
 \hline
& & & & & &  \\[-2ex]
 $\Phi_2$ &$ 1/2$ & $1/2$  & 
	$h_{33}^{(2)}/\sqrt{2}$  & $h_{32}^{(2)}/\sqrt{2}$ & 0 & 0
	\\
& & & & & &  \\[-2ex]
 \hline
& & & & & &  \\[-2ex]
 $\Phi_3$ &$ -5/2$ & $1/2$  & 
	$h_{33}^{(3)}/\sqrt{2}$  & $h_{32}^{(3)}/\sqrt{2}$ & 
  	$\tilde h_{33}^{(3)}$  & $\tilde h_{32}^{(3)}$
	\\
& & & & & &  \\[-2ex]
 \hline
& & & & & &  \\[-2ex]
 $\Phi_4$ &$ 1/2$ & $1/2$  & 
	$h_{33}^{(4)}/\sqrt{2}$  & $h_{32}^{(4)}/\sqrt{2}$ & 
        $\tilde h_{33}^{(4)}$  & $\tilde h_{32}^{(4)}$
	\\[1.5ex]
%& & & & & &  \\[-1ex]
 \hline
\end{tabular}
\end{center}
    \caption{\em Values of the coefficients $C_\Phi$, $C_t$, $l_b$, $l_s$, 
		$r_b$, $r_s$ for different types of Scalar Diquarks.
            }
	\label{tab:consts}
\end{table}

%% file: table.limits
\begin{table}[h]
\begin{center}
\begin{tabular}{|c|c|c|}
\hline
Products & \multicolumn{2}{c|}{Bounds from $B \rightarrow X_s \gamma$}
	\\ 
\cline{2-3}
 of Couplings & $ 1 \sigma$ & $ 2 \sigma$
	\\[0.9ex]
\hline
& & \\[-2ex]
 $h_{33}^{(1)} h_{32}^{(1)}$ &  
  	$[-1.0, -0.85 ] $ &
	        $ [-1.1, -0.74]$
        \\[0.9ex]
 \hspace*{-1em}or $h_{33}^{(3)} h_{32}^{(3)}$ 
  & 
     $[-9.3 \times 10^{-2},  5.8 \times 10^{-2}]$ & 
        	$[-0.2 , 0.12]$
	\\[0.9ex]
\hline
& & \\[-2ex]
$h_{33}^{(2)} h_{32}^{(2)}$ &  
	$[-0.16 , 0.25 ]$ &
		$ [-0.33, 0.53 ]$
	\\[0.9ex]
 \hspace*{-1em}or $h_{33}^{(4)} h_{32}^{(4)}$
	&       $[2.6, 3.0] $ & 
        	$[2.3, 3.2]$
  	\\[0.9ex]
\hline
& & \\[-2ex]
$h_{33}^{(3)} \tilde h_{32}^{(3)}$  & 
	$[-9.2 \times 10^{-4}, 9.2 \times 10^{-4}]$ & 
		$[-1.4 \times 10^{-3}, 1.4 \times 10^{-3}]$
	\\[0.9ex]
\hline
& & \\[-2ex]
$\tilde h_{33}^{(3)} h_{32}^{(3)}$ & 
	$[-2.2 \times 10^{-4}, 3.5 \times 10^{-4}]$ & 
		$[-4.5 \times 10^{-4}, 7.3 \times 10^{-4}]$
	\\[0.9ex]
& 	$ [3.3 \times 10^{-3}, 3.9 \times 10^{-3}]$ & 
	 	$[2.9 \times 10^{-3}, 4.1 \times 10^{-3}]$
	\\[0.9ex]
\hline
& & \\[-2ex]
$\tilde h_{33}^{(3)} \tilde h_{32}^{(3)}$  & 
	$[-0.12 , 0.12 ]$ & 
		$[-0.18, 0.18]$
	\\[0.9ex]
\hline
& & \\[-2ex]
$h_{33}^{(4)} \tilde h_{32}^{(4)}$  & 
	$[-2.5 \times 10^{-3}, 2.5 \times 10^{-3}]$  & 
		$[-3.7 \times 10^{-3}, 3.7 \times 10^{-3}]$
	\\[0.9ex]
\hline
& & \\[-2ex]
$\tilde h_{33}^{(4)} h_{32}^{(4)}$ & 
	$[-1.1 \times 10^{-2}, -9.3 \times 10^{-3}]$ & 
		$[-1.2 \times 10^{-2}, -8.3 \times 10^{-3}]$
	\\[0.9ex]
& 	$ [-9.4 \times 10^{-4}, 5.9 \times 10^{-4}]$ & 
		$[-2.0 \times 10^{-3}, 1.2 \times 10^{-3}]$
	\\[0.9ex]
\hline
& & \\[-2ex]
$\tilde h_{33}^{(4)} \tilde h_{32}^{(4)}$  & 
	$[-0.35 , 0.35 ]$ &
	 	$[-0.51, 0.51 ]$
	\\[1.1ex]
\hline
\end{tabular}
\end{center}
    \caption{\em Limits on the Scalar Diquark Couplings 
	     for $m_{\Phi_i}=100 ~GeV$.}
	\label{tab:limits}
\end{table}